\documentclass[twocolumn,showpacs,amsfonts,amsmath,amssymb,aps,pra,floatfix]{revtex4}
\usepackage{graphicx}
\usepackage{bm}

\begin{document}

\title{Harmonic oscillator in a rotating trap: Complete solution in 3D}
\author{Tomasz Sowi\'nski}
\email{tomsow@cft.edu.pl}\author{Iwo Bialynicki-Birula}
\email{birula@cft.edu.pl}\affiliation{Center for Theoretical Physics, Polish
Academy of Sciences\\ Al. Lotnik\'ow 32/46, 02-668 Warsaw, Poland and\\
Department of Physics, Warsaw University, Ho{\.z}a 69, 00-681 Warsaw, Poland}

\begin{abstract}
Complete description of the classical and quantum dynamics of a particle in an anisotropic, rotating, harmonic trap is given. The problem is studied in three dimensions and no restrictions on the geometry are imposed. In the generic case, for an arbitrary orientation of the rotation axis, there are two regions of instability with different characteristics. The analysis of the quantum-mechanical problem is made simple due to a direct connection between the classical mode vectors and the quantum-mechanical wave functions. This connection is obtained via the matrix Riccati equation that governs the time evolution of squeezed states of the harmonic oscillator. It is also shown that the inclusion of gravity leads to a resonant behavior --- the particles are expelled from the trap.
\end{abstract}

\pacs{03.75.Kk, 45.30.+s, 03.65.-w}

\maketitle

\section{Introduction}

Harmonic traps are often used in optics and atomic physics (especially in the study of Bose-Einstein condensates) and yet a complete theory of these devices has not been developed. In this work we present a complete solution to the problem of the motion of a single particle classical or quantum-mechanical moving in three dimensions for the most general anisotropic rotating harmonic trap. This is an exactly soluble problem but technical difficulties apparently served so far as a deterrent in developing a full description. All our results are valid also for the center of mass motion in many-body theory since for the quadratic potentials the center of mass motion completely separates for the internal motion \cite{kohn,dob,cmm}.

The solution to the problem of a two-dimensional rotating trap has been known for at least hundred years. In the classic textbook on analytical dynamics by Whittaker \cite{whitt} one finds a solution of a mathematically equivalent problem of small oscillations of ``a heavy particle about its position of equilibrium at the lowest point of a surface which is rotating with constant angular velocity about a vertical axis through the point''. Recently, this problem acquired a new significance in connection with the experimental and theoretical studies of Bose-Einstein condensates in rotating traps \cite{rzs,mad,sc,gg,ros}. Still, to our knowledge, a full description of the particle dynamics in a rotating anisotropic trap in three dimensions was not given. Explicit formulas describing the complete mode structure in the three-dimensional case are indeed quite cumbersome \cite{cmm} because we deal here with third-order polynomials and on top of that they have rather complicated coefficients. However, many important features may be exhibited without straining the reader's patience. In particular, we determine the characteristic frequencies for an arbitrary orientation of the angular velocity, and we identify various stability regions. Following a recent theoretical observation that the gravity may cause resonances \cite{ibb}, we give a full analysis of these phenomenon. Finally, owing to the properties of the matrix Riccati equations we unravel a direct connection between the solutions of the classical problem and the shape of the quantum wave functions. Also, we find the three constants of motion quadratic in the positions and momenta.

\section{Equations of motion}

The solution of our problem is the simplest in the co-rotating coordinate system, where the potential does not depend on time and the Hamiltonian has the following form:
\begin{equation}\label{ham}
 {\cal H} = \frac{\bm{p}^{2}}{2m}
 + \bm{r}\!\cdot\!\hat{\Omega}\!\cdot\!\bm{p}
 + \frac{m}{2}\bm{r}\!\cdot\!\hat{V}\!\cdot\!\bm{r}.
\end{equation}
The antisymmetric matrix $\hat{\Omega}$ is built from the components of the angular velocity vector ${\Omega}_{ik}=\epsilon_{ijk}\Omega_j$ and the potential matrix $\hat{V}$ is symmetric and positive definite. The term involving $\Omega$ is responsible for both the Coriolis force and the centrifugal force. In quantum mechanics the Hamiltonian (\ref{ham}) may be obtained by transforming the Schr\"odinger equation to the rotating frame by a substitution $\psi \to \exp(i{\bm\Omega}\!\cdot\!{\bm L} t/\hbar)\psi$.

The equations of motion determined by (\ref{ham}) look the same in classical and in quantum theory and they can be written in the following matrix form
\begin{equation}\label{eqnmot}
 \frac{dX(t)}{dt} = \hat{M}X(t),
\end{equation}
where the six-dimensional vector $X(t)$ comprises coordinates and momenta $X=(x,y,z,p_x,p_y,p_z)$ and the $6\times 6$ matrix $\hat{M}$ is \begin{eqnarray}\label{defm}
 \hat{M} = \begin{pmatrix}
  0 & \Omega_z & -\Omega_y & 1/m & 0 & 0\\
  -\Omega_z & 0 & \Omega_x & 0 & 1/m & 0\\
  \Omega_y & -\Omega_x & 0 & 0 & 0 & 1/m\\
 - mV_{xx} & -mV_{xy} & -mV_{xz} & 0 & \Omega_z & -\Omega_y\\
 - mV_{yx} & -mV_{yy} & -mV_{yz} & -\Omega_z & 0 & \Omega_x\\
 - mV_{zx} & -mV_{zy} & -mV_{zz} & \Omega_y & -\Omega_x & 0\\
\end{pmatrix}\!\!.
\end{eqnarray}
The set of six linear differential equations (\ref{eqnmot}) has in general six independent solutions $X_i(t)$ (normal modes) that in the stability region change harmonically with time. The frequencies of the normal modes are determined by the characteristic polynomial $P(\omega)$ associated with the matrix $\hat{M}$. Owing to the time-reversal symmetry of our problem, the characteristic polynomial is triquadratic
\begin{equation}\label{charpol}
 P(\omega) = \omega^6 + A\,\omega^4 + B\,\omega^2 + C.
\end{equation}
The coefficients $A,B,C$ have been obtained in our previous publication \cite{cmm} and have the following rotationally invariant forms
\begin{subequations}
\begin{eqnarray}
A &=&-2\Omega^2 - \textrm{Tr}\{\hat{V}\},\\
B &=&\!\Omega^4\!+\!\Omega^2(3\bm{n}\!\cdot\!\hat{V}\!\cdot\!\bm{n}\!
-\!\textrm{Tr}\{\hat{V}\})\!
+\!\frac{\textrm{Tr}\{\hat{V}\}^2\!-\!\textrm{Tr}\{\hat{V}^2\}}{2},\\
C &=&{\Omega}^2(\textrm{Tr}\{\hat{V}\}\!
-\!{\Omega}^2)\bm{n}\!\cdot\!\hat{V}\!\cdot\!\bm{n}\!
-\!\Omega^2\bm{n}\!\cdot\!\hat{V^2}\!\cdot\!\bm{n}\!
-\!\textrm{Det}\{\hat{V}\},\qquad
\end{eqnarray}
\end{subequations}
where $\Omega$ is the length and $\bm{n}$ is the direction of the angular velocity vector, $\bm{\Omega}=\Omega \bm{n}$. Since the matrix $\hat{V}$ is by definition symmetric, one may always choose a base in the rotating frame such that $\hat{V}$ will be diagonal ($\hat{V}=\textrm{diag}(V_x,V_y,V_z)$). As was to be expected, the mass has dropped out completely from the characteristic equation. Therefore, all the results will scale uniformly and one may express them in terms of an arbitrarily chosen frequency unit $\omega_0$. In what follows, all the dimensionless frequencies are to be the multiplied by $\omega_0$ and the trap parameters $V_i$ by $\omega_0^2$.

\section{Stability of motion}

\begin{figure}
\centering
\includegraphics[angle=-90,width=0.45\textwidth]{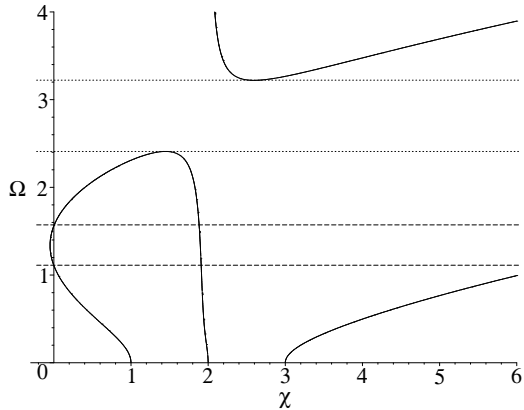}
\caption{The contour lines in this figure represent the zeroes of the characteristic polynomial (\ref{charpol1}) for $V_x=1$, $V_y=2$, $V_z=3$ and $n_x=1/\sqrt{3}$, $n_y=1/\sqrt{3}$, $n_z=1/\sqrt{3}$ plotted as functions the magnitude of angular velocity $\Omega$ and the square $\chi=\omega^2$ of the characteristic frequency. In addition to the lower region of instability (between dashed lines) where one of the roots of (\ref{charpol1}) is negative, there is also the upper region (between dotted lines) where there exists only one real root (described by the continuation of the line that begins at $\chi=3$, not seen in this figure because it corresponds to a very large value of $\chi$).}\label{fig1}
\end{figure}

The stability of motion for a harmonic oscillator is determined by the values of its characteristic frequencies $\omega$ --- the roots of the characteristic polynomial. Stable oscillations take place when all $\omega$'s are real. This means that all three roots of the polynomial $Q(\chi)$
\begin{equation}\label{charpol1}
Q(\chi)=\chi^3 + A\,\chi^2 + B\,\chi + C,\;\;\chi=\omega^2
\end{equation}
are real and positive. It has been argued in Ref.~\cite{cmm} that there is always a region of instability; this occurs when one of the roots of $Q(\chi)$ is negative. The corresponding modes grow exponentially with time. As seen in Fig.~\ref{fig1}, this region of instability is bounded by the two values $\Omega_\pm$ at which the curve crosses the vertical axis. These values are given by the zeroes of $C$, treated as a biquadratic expressions in $\Omega$
\begin{equation}\label{zeroes}
\Omega_\pm = \sqrt{\frac{b\pm\sqrt{b^2-4ac}}{2a}},
\end{equation}
where $a=\bm{n}\!\cdot\!\hat{V}\!\cdot\!\bm{n},\;
b=\textrm{Tr}\{\hat{V}\}\bm{n}\!\cdot\!\hat{V}\!\cdot\!\bm{n}
-\bm{n}\!\cdot\!\hat{V^2}\!\cdot\!\bm{n}$, and $c=\textrm{Det}\{\hat{V}\}$. Since $a, b$, and $c$ are positive, $\Omega_\pm$ are real.

In this paper we show that a more detailed investigation reveals that in addition to the region of (purely exponential) instability when one root of $Q(\chi)$ is negative there is also, in general, an additional region of (oscillatory) instability where two roots are complex. Without rotation, when $\Omega=0$, the three roots of $Q(\chi)$ are equal to the eigenvalues of the potential matrix $\hat{V}$ and one has a simple system of three harmonic oscillators vibrating independently along the principal directions of the trap. As $\Omega$ increases, our system has, in general, two regions of instability: the lower one when one of the roots of $Q(\chi)$ is negative and the upper one when two roots are complex. We shall exhibit this behavior by plotting the zero contour lines of (\ref{charpol1}) in the $\Omega\chi$-plane. We assume that the trap potential and the direction of rotation are fixed and we treat the characteristic polynomial $Q(\chi)$ as a function of $\Omega$ and $\chi$ only. Contour lines representing the zeroes of (\ref{charpol1}) in the generic case are shown in Fig.~\ref{fig1}. There is a region of $\Omega$, where only one real solution of (\ref{charpol}) exists. However, this region is bounded, so for sufficiently large $\Omega$ the system is always stable.

\begin{figure}
\centering
\includegraphics[angle=-90,width=0.45\textwidth]{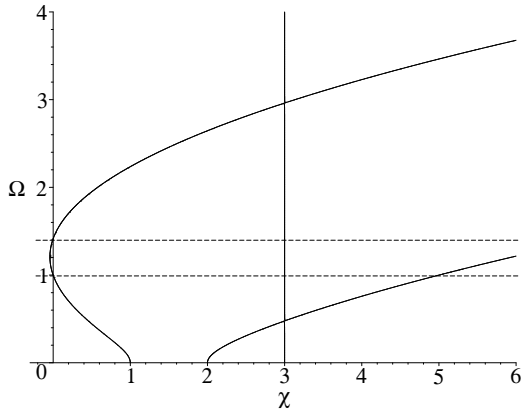}
\caption{This plot is for the same trap as in Fig.~\ref{fig1} but for the rotation around the trap axis $n_x=0$, $n_y=0$, $n_z=1$. In this degenerate case there exists only one region of instability, when one of the roots of (\ref{charpol1}) is negative.} \label{fig2}
\end{figure}
\begin{figure}
\centering
\includegraphics[angle=-90,width=0.45\textwidth]{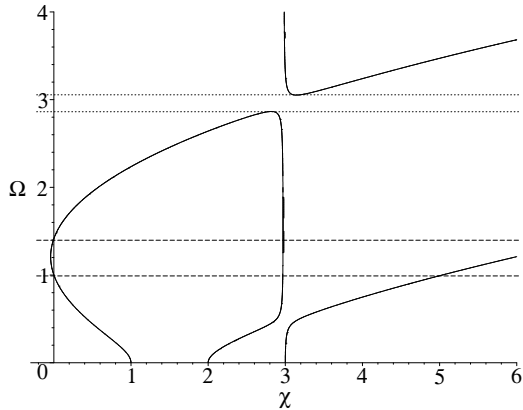}
\caption{This plot shows how the degenerate case (Fig.~\ref{fig2}) merges with the general case (Fig.~\ref{fig1}). Here we have chosen the following values of the parameters: $V_x=1$, $V_y=2$, $V_z=3$ and $n_x=\sin(1/10)$, $n_y=0$, $n_z=\cos(1/10)$.} \label{fig3}
\end{figure}

Graphical representation of the solutions for $n_x=1$, $n_y=n_z=0$ is shown in Fig.~\ref{fig2}. In this plot only one instability region is seen, where one of the roots of (\ref{charpol1}) is negative. Owing to the stabilizing effect of the Coriolis force, for fast rotations the system becomes again stable. In this plot one can also see, that one of the characteristic frequencies remains constant. It is so, because the rotation does not influence the motion in the direction of the rotation axis. When $\Omega$ increases, the solutions of (\ref{charpol1}) change their character. In the previous case there was only one instability region, where the square of the frequency was negative. When the direction of angular velocity is not parallel to one of the axes of the trap, a second kind of instability appears --- the square of the frequency becomes complex. Since the coefficients of the characteristic polynomial are real, the square of the second frequency is complex conjugate to the first one. In this case the instability has the form of expanding oscillations. In Fig.~\ref{fig3} we show what happens when the rotation axis is tilted away a little bit from the $z$ direction and the second kind of instability develops. It has been shown in \cite{cmm} that when the trap is not fully symmetric the first kind of instability exist for every direction of the angular velocity.

We will show now, that the second kind of instability does not occur when the rotation vector is along one of the axes of the trap. Therefore, it can never happen in a two dimensional trap.
Without any loss of generality, one may assume that the direction of rotation is along the $z$ axis. In this degenerate case the characteristic polynomial (\ref{charpol}) factorizes because the motion in the $z$ direction is not influenced by rotation,
\begin{eqnarray}
\label{factor}
(\chi^2-\chi(2\Omega^2+V_x+V_y)+\Omega^4
-\Omega^2(V_x+V_y)+V_x V_y)\nonumber\\
\times(\chi - V_z)=0.\;\;\;
\end{eqnarray}

The biquadratic polynomial in (\ref{factor}) has real zeros if its discriminant $\Delta$ is positive and this is, indeed, the case since \begin{eqnarray}
\label{delta}
\Delta &=& (2\Omega^2+V_x+V_y)^2-4(\Omega^4-\Omega^2 (V_x+V_y)+V_x V_y)\nonumber \\
&=& 8\Omega^2(V_x+V_y)+(V_x-V_y)^2 \geq 0.
\end{eqnarray}

The second kind of instability always exists if the direction of rotation does not coincide with one of the axes of the harmonic trap.  The tilting the axis of rotation away from the trap axis not only leads to the appearance of an additional region of instability but also, as shown in the next Section, leads to new effects when gravity is taken into account.

\section{Role of gravity}

So far, we have neglected the presence of the gravitational field. It turns out that, owing to a possible resonance \cite{ibb}, gravity may strongly influence the center of mass motion and lead to an instability. In the frame co-rotating with the trap the gravitational field is seen as a time dependent periodic force. This force, under appropriate conditions, may fall into resonance with one of the characteristic frequencies of the trap and cause the linear growth of the amplitude and the particles will escape from the trap. Of course, such a homogeneous external field will only influence the center of mass motion; it will not change the dynamics of the internal, relative motion. This is all true in the classical and in the quantum theory.

\begin{figure}
\centering
\includegraphics[angle=-90,width=0.45\textwidth]{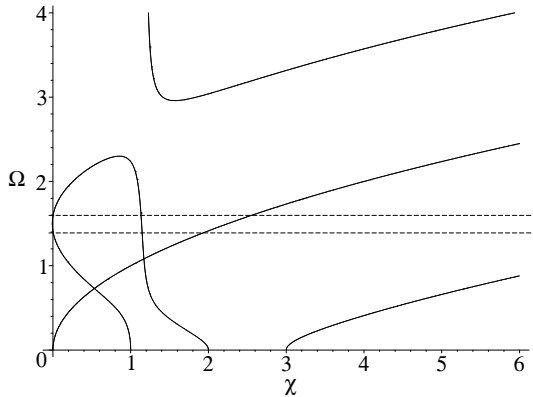}
\caption{$V_x=1$, $V_y=2$, $V_z=3$ $n_x=\sin(\frac{2\pi}{5})$, $n_y=0$, $n_z=\cos(\frac{2\pi}{5})$. Two horizontal lines enclose the lower region of instability. Both resonant frequencies lie in the lower region of stability.} \label{fig_rez1}
\end{figure}

\begin{figure}
\centering
\includegraphics[angle=-90,width=0.45\textwidth]{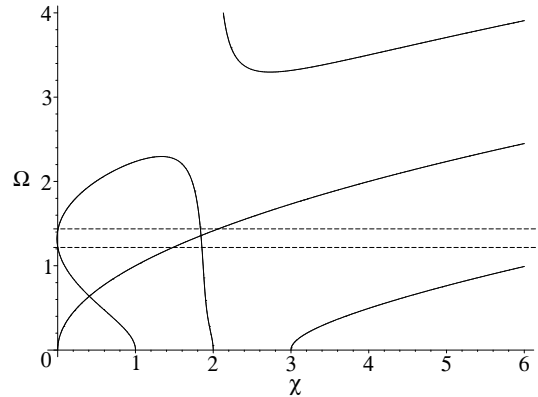}
\caption{$V_x=1$, $V_y=2$, $V_z=3$ $n_x=\sin(\frac{\pi}{4})$, $n_y=0$, $n_z=\cos(\frac{\pi}{4})$. Higher resonant frequency lies in the lower region of instability.} \label{fig_rez2}
\end{figure}

\begin{figure}
\centering
\includegraphics[angle=-90,width=0.45\textwidth]{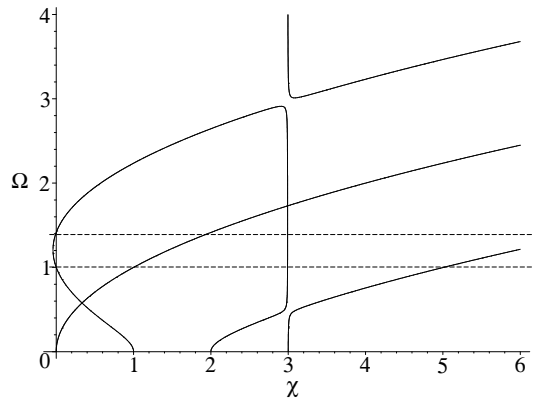}
\caption{$V_x=1$, $V_y=2$, $V_z=3$ $n_x=\sin(\frac{\pi}{60})$, $n_y=0$, $n_z=\cos(\frac{\pi}{60})$. Resonant frequencies lie in two different regions of stability.} \label{fig_rez3}
\end{figure}

The equations of motion for the center of mass in the presence of gravity acquire an additional term as compared to (\ref{eqnmot})
\begin{equation}
 \frac{dX(t)}{dt} = \hat{M}X(t)+ \begin{pmatrix} 0 \\ m\bm{g}(t) \end{pmatrix}.\\
\end{equation}
In the rotating frame the gravitational acceleration can be written in the form \cite{ibb}
\begin{equation}
\bm{g}(t) = \Re\left(\bm{g}_{\parallel}
+(\bm{g}_{\perp}+i(\bm{n}\times\bm{g}_{\perp}))\mathrm{e}^{i\Omega t}\right),
\end{equation}
where $\bm{g}_{\parallel}$ and $\bm{g}_{\perp}$ denote the parts of the gravitational acceleration parallel and orthogonal at $t=0$ to the direction of rotation $\bm{n}$. Thus, we have arrived at the classic problem of a periodically forced harmonic oscillator in three dimensions. Resonances will occur when the frequency of the external force $\Omega$ coincides with one of the three characteristic frequencies $\omega_i$ of the oscillator. An interesting novelty here, as compared to the standard forced oscillator, is that the characteristic frequencies themselves depend on $\Omega$. In order to find the condition for a resonance, one has to put $\omega=\Omega$ in (\ref{charpol}) and solve  for $\Omega$ the resulting biquadratic equation
\begin{equation} \label{reseq}
 D \Omega^4 + E \Omega^2 +F = 0,
\end{equation}
where
\begin{eqnarray}
D &=& -2(\,\mathrm{Tr}\{\hat{V}\}-\bm{n}\!\cdot\!\hat{V}\!\cdot\!\bm{n}\,), \nonumber \\
E &=& \left(\!\frac{\textrm{Tr}\{\hat{V}\}^2\!-\!\textrm{Tr}\{\hat{V}^2\}}{2}+\textrm{Tr}\{\hat{V}\}\bm{n}\!\cdot\!\hat{V}\!\cdot\!\bm{n}\! - \bm{n}\!\cdot\!\hat{V^2}\!\cdot\!\bm{n}\! \right),\nonumber \\
F &=& -\mathrm{Det}\{\hat{V}\}.
\end{eqnarray}
As was shown in \cite{ibb} this equation always has two real solutions for $\Omega^2$, the smaller root lies in the lower stability region. It turns out that we can choose the parameters of the trap in such a way that the larger root falls into the lower region of stability (Fig.~\ref{fig_rez1}), into the lower region of instability (Fig.~\ref{fig_rez2}), or into the higher region of stability (Fig.~\ref{fig_rez3}). These figures also illustrate the simplest graphical method to find the resonant frequencies. Since $\chi = \omega^2$, in our figures \ref{fig1}-\ref{fig3} that depict the locations of the stability regions, the positions of resonances are found at the intersections of the parabola $\chi=\Omega^2$ with the previously drawn curves.

Full analytical description is very simple in the two-dimensional case, when the trap rotates around one of its principal axes. For definitness, let us assume that the $\Omega$ is along the $z$ axis whose orientation in space is still arbitrary. Without any loss of generality, one may assume that $V_x<V_y$. In this case, Eq.~(\ref{reseq}) has two simple solutions.
\begin{equation}
\Omega_1^2=\frac{V_x V_y}{2(V_x + V_y)}, \qquad \Omega_2^2=V_z.
\end{equation}
From Eq.~(\ref{zeroes}) one can deduce that the lower region of stability lies below $\Omega^2=V_x$, the region of instability is in the range $V_x  \leq \Omega^2 \leq V_y$, and the higher region of stability extends above $\Omega^2 = V_y$. One can also see that $\Omega_1^2<V_x$. Thus, the lower resonant frequency always falls into the first region of stability. The location of the higher resonant frequency varies. When  $V_z<V_x$, then both frequencies lie in the lower region of stability.  $V_z>V_y$ then the higher resonant frequency falls into the higher region of stability. In the last case, when $V_x  \leq V_z \leq V_y$ the higher resonant frequency lies in the region of instability. This case in not interesting because in this region the amplitude grows exponentially anyway and the resonance will not be noticeable. In the special case of the rotation around one of the trap axis, the position of resonant frequencies does not depend on the orientation of the rotation axis with respect to the vertical direction. In the general case, the calculations are much more complicated.

\section{Classical-Quantum correspondence}
\subsection{Matrix Riccati equation}
Let us consider a state of a quantum-mechanical system with the Hamiltonian (\ref{ham}) described by a Gaussian wave function of the form:
\begin{equation}\label{ffal}
\Psi = C(t)\exp(-\frac{1}{2}{\bm r}\!\cdot\!\hat{K}(t)\!\cdot\!{\bm r}),
\end{equation}
where the matrix $\hat{K}$ is of course symmetric (its real part is positive definite) and $C$ is a normalization constant,
\begin{equation}
C(t) = \sqrt{{\rm Det}\{\Re(\hat{K}(t))/\sqrt{\pi}\}}.
\end{equation}
From the Schr\"odinger equation $i\hbar\partial_t \Psi = {\cal H} \Psi$ with the Hamiltonian (\ref{ham}) one obtains the evolution equation for the matrix $\hat{K}$ in the form of a matrix Riccati equation ($\hbar=1, m=1$)
\begin{equation}\label{riccati}
\frac{d\hat{K}}{dt} = -i\hat{K}^2+i\hat{V} - [\hat{\Omega},\hat{K}].
\end{equation}
Mathematical theory of these equations is well developed \cite{reid,zwilling}. In particular, the matrix Riccati equations find numerous applications in control theory \cite{ggs}. It is known that these nonlinear equations can be replaced by the linear ones. To this end we shall follow the procedure outlined by us previously \cite{beyond} in the same quantum-mechanical context and we shall search for solutions of Eq.~(\ref{riccati}) in the form
\begin{equation}\label{decomp}
\hat{K}(t) = -i \hat{N}(t)\hat{D}^{-1}(t),
\end{equation}
which leads to
\begin{eqnarray}
-i\frac{d\hat{N}}{dt}\hat{D}^{-1}
+i\hat{N}\hat{D}^{-1}\frac{d\hat{D}}{dt}\hat{D}^{-1}& \nonumber\\
=i\hat{N}\hat{D}^{-1}\hat{N}\hat{D}^{-1}
+i\hat{V}+i\hat{\Omega}\hat{N}\hat{D}^{-1}
-\!&i\hat{N}\hat{D}^{-1}\hat{\Omega}.
\end{eqnarray}
This matrix equation is satisfied when the matrices $\hat{N}$ and $\hat{D}$ obey the following {\em linear equations}
\begin{subequations}\label{eqmot2}
\begin{eqnarray}
\frac{d\hat{D}}{dt}&=& \hat{N} - \hat{\Omega}\hat{D},\\
\frac{d\hat{N}}{dt}&=& -\hat{V}\hat{D} - \hat{\Omega}\hat{N}.
\end{eqnarray}
\end{subequations}
The linearization of the Riccati equation, in addition to being an effective mathematical tool, has also conceptual advantages. Namely, it leads to a direct relationship between classical and quantum theory. Comparing (\ref{eqmot2}) with (\ref{eqnmot}), one can see that the columns of the matrices $\hat{D}$ and $\hat{N}$ satisfy the same equations as the classical position and momentum vectors, respectively. Therefore, from the knowledge of the classical motion, one may determine the evolution of a Gaussian wave function. This relationship between classical and quantum mechanics has been noticed before by Arnaud \cite{arnaud} but only in the special case of the one-dimensional oscillator, when there is no nontrivial mode structure. Here, we extend it to three dimensions and we also give a prescription how to construct {\em stationary} quantum states from the classical mode vectors.

\subsection{Stationary quantum states}\label{b}

In a stationary quantum state (\ref{ffal}) the matrix $\hat{N}(t)$ must have the following form:
\begin{equation}
\label{indep} \hat{N}(t) = i\hat{K}_0\hat{D}(t),
\end{equation}
where $\hat{K}_0$ does not depend on time. Inserting (\ref{indep}) into (\ref{eqmot2}) gives an obvious relation (cf. Eq.~{\ref{riccati}})
\begin{eqnarray}
\label{riccati1} 0 = -i\hat{K}_0^2 + i\hat{V} - [\hat{\Omega},\hat{K}_0].
\end{eqnarray}
However, we gained a new insight: stationary solutions of the Riccati equation (\ref{riccati}) can be constructed from the solutions of the set of linear equations (\ref{eqnmot}) of classical mechanics.

In order to construct a stationary solution of (\ref{riccati}) for a harmonic oscillator in $d$ dimensions we need $d$ linearly independent eigenmodes of the classical equations of motion. Such eigenmodes evolve in time according to the formulas $X_i(t)={\bar X}_i\exp(i\omega_i t)$. Let us construct now two $d\times d$ matrices ${\hat x}(t)$ and ${\hat p}(t)$ whose columns represent the positions and momenta, respectively. These matrices can be written in the form
\begin{eqnarray}
\label{matrices} {\hat x}(t) = {\hat x}_0\exp(i{\hat\omega}t),\;\;{\hat
p}(t) = {\hat p}_0\exp(i{\hat\omega}t),
\end{eqnarray}
where ${\hat\omega}$ is a diagonal matrix whose elements are equal to the frequencies $\omega_i$ of the chosen modes. Since the matrices ${\hat x}(t)$ and ${\hat p}(t)$ are solutions of Eqs.~(\ref{eqmot2}), one may identify them with the matrices ${\hat N}(t)$ and ${\hat D}(t)$. The matrix ${\hat K}={\hat x}(t){\hat p}(t)^{-1}$ will satisfy the matrix Riccati equation. Moreover, owing to the cancellation of the time dependent terms $\exp(i{\hat\omega}t)$, it will be a time-independent solution.

In addition to the special stationary solution, there are also many nonstationary solutions of Eq.~(\ref{riccati}). They describe pulsating Gaussians and they are the analogs of squeezed states, known from quantum optics.

\subsection{Two-dimensional harmonic oscillator}

In general, Eq.~(\ref{riccati1}) for $\hat{K}_0$ leads to a system of three quadratic equations that are hard to solve but one can find the solutions quite easily in the simple case --- for a the two-dimensional rotating harmonic trap (the Whittaker problem). This explicitly soluble example will be used to explain the connection between the classical and quantum cases.

We shall choose the coordinate axes along the directions of the trap to make the potential matrix diagonal $\hat{V}=\textrm{diag}(V_x,V_y,V_z)$. The angular rotation vector is parallel to $z$ axis of the potential ellipsoid. We shall parametrize the solutions of Eq.~(\ref{riccati1}) as follows
\begin{equation}
\label{fink}
\hat{K}_0 = \begin{pmatrix}
\alpha & i\gamma & 0\\
i\gamma & \beta & 0 \\
0 & 0 & \sqrt{V_z}
\end{pmatrix}.\\
\end{equation}
and obtain the following set of algebraic equations for $\alpha, \beta$, and $\gamma$
\begin{subequations}
\label{3eqns}
\begin{eqnarray}
\gamma^2-\alpha^2+2\Omega \gamma +V_x=0,\label{3aeqns}\\
\gamma^2-\beta^2-2\Omega \gamma +V_y=0,\label{3beqns}\\
\alpha(\Omega-\gamma)=\beta(\Omega+\gamma).\label{3ceqns}
\end{eqnarray}
\end{subequations}
The solution of these equations proceeds in three simple steps. We first solve Eqs.~(\ref{3aeqns}) and (\ref{3beqns}) for $\alpha$ and $\beta$ choosing the positive square roots to secure positivity of the real part of $\hat{K}$
\begin{subequations}
\label{2solns}
\begin{eqnarray}
\alpha=\sqrt{V_x-\Omega^2+(\gamma+\Omega)^2},\label{2asolns}\\
\beta=\sqrt{V_y-\Omega^2+(\gamma-\Omega)^2}.\label{2bsolns}
\end{eqnarray}
\end{subequations}
Next we substitute these expressions into Eq.~(\ref{3ceqns}) to find $\gamma$ from the quadratic equation that results from squaring both sides of (\ref{3ceqns}). Only one solution of this quadratic equation satisfies (\ref{3ceqns}) and it can be written in the form
\begin{eqnarray}
\gamma=\Omega\frac{\kappa + 1}{\kappa - 1},\label{cc}
\end{eqnarray}
where $\kappa=\pm\sqrt{(V_x-\Omega^2)/(V_y-\Omega^2)}$ is a real parameter in both stability regions. In the first region the sign of $\kappa$ is positive, while in the second region it is negative. In the last step, we substitute the value (\ref{cc}) of $\gamma$ into Eqs.~(\ref{2solns}) to obtain the final formulas for the parameters $\alpha$ and $\beta$
\begin{subequations}
\label{1solns}
\begin{eqnarray}
\alpha=\sqrt{V_x-\Omega^2+4\Omega^2/(1-1/\kappa)^2},\label{1asolns}\\
\beta=\sqrt{V_y-\Omega^2+4\Omega^2/(1-\kappa)^2},\label{1bsolns}
\end{eqnarray}
\end{subequations}
The matrix (\ref{fink})
is symmetric, positive definite in all stability regions and it satisfies the equation (\ref{riccati1}) for stationary states.

In the three dimensional case a direct solution of the nonlinear equations for $\hat{K}_0$ is impossible since it requires solving high order polynomial equations. However, the method that employs a representation of $\hat{K}$ in the form (\ref{decomp}) requires only the solution of linear equations.

\subsection{Construction in terms of the classical modes}

We shall show now that the general procedure described in \ref{b} reproduces in two dimensions the solution (\ref{fink}). In the two-dimensional case there are four solutions corresponding to four characteristic frequencies $\pm\omega_+,\pm\omega_-$:
\begin{eqnarray*}
\omega_+^2 = \frac{V_x+V_y+2\Omega^2
+\sqrt{(V_x-V_y)^2+8\Omega^2(V_x+V_y)}}{2},\\
\omega_-^2 = \frac{V_x+V_y+2\Omega^2
-\sqrt{(V_x-V_y)^2+8\Omega^2(V_x+V_y)}}{2},\\
X(\omega)=\begin{pmatrix} x(\omega) \\ y(\omega) \\ p_x(\omega)\\
p_y(\omega)\end{pmatrix} = N
\begin{pmatrix} 2i\omega\Omega \\ V_x - \omega^2 - \Omega^2 \\
\Omega (\Omega^2-\omega^2-V_x) \\
i\omega(\Omega^2-\omega^2+V_x)
\end{pmatrix},
\end{eqnarray*}
where $N$ is a normalization constant and $\omega$ is one of the four characteristic frequencies.

From any two modes one may construct the following matrices:
\begin{subequations}
\label{prescr}
\begin{eqnarray}
\hat{\cal{D}} &=& \begin{pmatrix} x(\omega_1) & x(\omega_2)\\
 y(\omega_1) & y(\omega_2)\end{pmatrix},\\
\hat{\cal{N}} &=& \begin{pmatrix} p_x(\omega_1) & p_x(\omega_2)\\
p_y(\omega_1) & p_y(\omega_2)\end{pmatrix},
\end{eqnarray}
\end{subequations}
and then construct $\hat{K}_0=-i\hat{\cal{N}}\hat{\cal{D}}^{-1}$. This matrix will satisfy the relation (\ref{riccati1}) for a stationary solution of the Schr\"odinger equation. There is still the condition that the real part of $\hat{K}_0$ is positive definite to guarantee that the solution is square integrable. One may check that the choice of two modes that differ only by the sign of the characteristic frequency does not give the right solution since in this case the real part of $\hat{K}$ has one eigenvalue zero. This leaves us with four possible choices of the signs of two different frequencies. When $\Omega$ lies in the first stability region ($\Omega<\min(V_x,V_y)$) then both signs must be chosen positive. When $\Omega$ lies in the second stability region ($\Omega>\max(V_x,V_y)$) then one must choose the opposite signs with the minus signs accompanying the lower frequency $\omega$. In the region of instability it is not possible to choose the signs in such a way that the real part of the matrix $\hat{K}$ is positive definite. The matrix $\hat{K}$ constructed in this way is unique (the normalization constants $N$ cancel out) and coincides with the previously calculated matrix (\ref{fink}). The choice of signs is determined by the diagonal form of the classical Hamiltonian:
\begin{equation}\label{ham1}
{\cal H} = \omega_1a^*_1a_1 + \omega_2a^*_2a_2,
\end{equation}
where the complex amplitudes $a_i$ and $a^*_j$ satisfy the following Poisson bracket relations:
\begin{equation}
\{a_i,a^*_j\}=-i\delta_{ij} \qquad \{a_i,a_j\}=\{a^*_i,a^*_j\}=0.
\end{equation}
In the first stability region both terms in (\ref{ham1}) are positive while in the second region the term corresponding to smaller value of $\omega$ is negative. This is exactly the prescription used in constructing the solution (\ref{prescr}).

Having found the wave functions, one may proceed to calculate the Wigner functions. They are especially useful for a comparison with classical physics. For Gaussian wave functions, the Wigner function is an exponential of a quadratic form of positions and momenta
\begin{equation}
W(\bm{r},\bm{p})=M \exp(-\frac{1}{2}X\!\cdot\!\hat{W}\!\cdot\!X).
\end{equation}
For stationary states this quadratic form is a constant of motion and as such must be a combination of the expressions (\ref{const}). In the two-dimensional case it simply reduces to:
\begin{equation}
W(\bm{r},\bm{p})=M \exp(k_1{\cal C}_1 + k_2 {\cal C}_2),
\end{equation}
where ${\cal C}_1$ and ${\cal C}_2$ are the constants of motion defined by (\ref{SS2d}) and
\begin{eqnarray}
k_1 &=& \frac{2(\beta V_y-\alpha V_x)}{\alpha \beta (V_y-V_x)}, \\
k_2 &=& \frac{2(\alpha - \beta)}{\alpha \beta (V_y - V_x)}.
\end{eqnarray}

The construction of the matrix $\hat{K}$ from the solutions of the classical equations of motions in three dimensions proceeds along similar lines. In this case the solution of the nonlinear equations (\ref{riccati1}) by a direct method is more complicated, if possible at all. Our method of constructing $\hat{K}$ from the eigenmodes is based on the following unambiguous prescription, how to choose from the six modes the appropriate three modes. First, the three modes must have different absolute values of the frequency. The choice of the signs depends on the region of stability. In the first region, all signs are positive, in the second the smallest frequency must be taken with the negative sign, while in the third region it is the middle frequency.

It is clear from our construction of the stationary Gaussian solution of the Schr\"odinger equation from classical modes shows that the classical and the quantum problems always share the same regions of stability and instability.

\section{Conclusions}

We have presented a full analysis of the dynamical properties of an anisotropic, rotating harmonic trap. The main result of this analysis is the discovery of three distinct regions of stability and two regions of instability. In the generic case, by increasing the rate of rotation we visit successively all these regions. We have also shown that, owing to a direct link between the solutions of the Newton equations and the Schr\"odinger equation, the same regions of stability/instability are characteristic of both the classical and the quantum case. The dynamical properties of rotating traps are further enriched by the presence of gravity if the axis of rotation is not vertical. Gravity-induced resonances cause, at sharply defined rotation rates, an escape of particles from the trap.

This research was supported by a grant from the Polish Committee for Scientific Research in the years 2004-06.

\appendix

\section{Constants of motion}

The Hamiltonian of every harmonic oscillator in three dimensions can be written as a sum of three independent terms
\begin{equation}\label{ham2}
{\cal H} = \omega_1a^\dagger_1a_1 + \omega_2a^\dagger_2a_2 +
\omega_3a^\dagger_3a_3,
\end{equation}
where the coefficients $\omega_k$ are not necessarily positive. In the classical theory $a$'s and $a^\dagger$'s are complex amplitudes while in the quantum case they become annihilation and creation operators. It is clear that each part of (\ref{ham2}) is a constant of motion. These constants expressed in terms of ${\bm p}$ and ${\bm r}$ have the general form
\begin{equation}
\label{const} {\cal C}=\frac{1}{2}{\bm p}\!\cdot\!\hat{T}\!\cdot\!{\bm p} +
{\bm r}\!\cdot\!\hat{W}\!\cdot\!{\bm p} +\frac{1}{2}{\bm r}\!\cdot\!\hat{U}\!\cdot\!{\bm r}
\end{equation}
where the matrices $\hat{T},\hat{W},\hat{U}$ must have a rotationally invariant representation in terms of $\bm{\Omega}$ and $\hat{V}$. The time evolution of $\cal C$, as determined by the equations of motion (\ref{eqnmot}), leads to the following equations for $\hat{T},\hat{U},\hat{W}$
\begin{equation}
\label{eqs} \left\{ \begin{array}{lcl} 0 = \dot{\hat{T}} &=&
  [\hat{\Omega},\hat{T}]+\hat{W}+ \hat{W}^T\\
0 = \dot{\hat{U}} &=&
  [\hat{\Omega},\hat{U}]-\hat{W}\hat{V}- \hat{V}\hat{W}^T \\
0 = \dot{\hat{W}} &=&
  [\hat{\Omega},\hat{W}]+\hat{U}- \hat{V}\hat{T}
\end{array} \right.
\end{equation}
The number of independent quadratic constants of motion is the same as the number of dimensions of the harmonic oscillator. They can be chosen as the bilinear combinations of the mode amplitudes $a^*_k a_k$.

In two dimensions we have
\begin{subequations}
\label{SS2d}
\begin{eqnarray}
\label{S12d} &&{\cal C}_1 = \left\{ \begin{array}{rcl}
\hat{T}&=& \hat{\openone}\\
\hat{W}&=& \hat{\Omega}\\
\hat{U}&=& \hat{V}
\end{array} \right.\\
\label{S22d} &&{\cal C}_2 = \left\{\begin{array}{rcl}
\hat{T}&=& \hat{V}\\
\hat{W}&=& \hat{\Omega}\hat{V}+2\hat{V}\hat{\Omega} +2\hat{\Omega}^3 \\
\hat{U}&=& \hat{V}^2 + \hat{V}\hat{\Omega}^2 - \hat{\Omega}\hat{V}\hat{\Omega}
\end{array} \right.
\end{eqnarray}
\end{subequations}

In three dimensions there are three independent solutions of Eqs.~(\ref{eqs})
\begin{widetext}
\begin{subequations}
\label{SS}
\begin{eqnarray}
\label{S1} &&{\cal C}_1 = \left\{ \begin{array}{rcl}
\hat{T}&=&\hat{\openone}\\
\hat{U}&=& \hat{V}\\
\hat{W}&=& \hat{\Omega}
\end{array} \right.\\
\label{S2} &&{\cal C}_2 = \left\{\begin{array}{rcl}
\hat{T}&=& \hat{V}-3\hat{\Omega}^2\\
\hat{U}&=& \hat{V}^2-\hat{V}\hat{\Omega}^2-\hat{\Omega}^2\hat{V}
-\hat{\Omega}\hat{V}\hat{\Omega}\\
\hat{W}&=& \hat{\Omega}\hat{V}+2\hat{V}\hat{\Omega}-\hat{\Omega}^3
\end{array} \right.\\
\label{S3} &&{\cal C}_3 = \left\{\begin{array}{rcl} \hat{T}&=&3\hat{V}^2 +4\hat{\Omega}^2\hat{V}+4\hat{V}\hat{\Omega}^2
+\hat{\Omega}\hat{V}\hat{\Omega}+8\Omega^2\hat{V}
-13\textrm{Tr}\{\hat{V}\}(\Omega^2\hat{\openone}-\hat{\Omega}^2)\\
\hat{U}&=& 3\hat{V}^3-2\hat{V}\hat{\Omega}^4-2\hat{\Omega}^4\hat{V}
+3\hat{\Omega}^2\hat{V}\hat{\Omega}^2-\Omega^2\hat{\Omega}\hat{V}\hat{\Omega}
-3\hat{V}^2\hat{\Omega}^2-3\hat{\Omega}^2\hat{V}^2
-3\hat{\Omega}\hat{V}^2\hat{\Omega}\\
&&-9\hat{V}\hat{\Omega}^2\hat{V} -6\hat{V}\hat{\Omega}\hat{V}\hat{\Omega}
-6\hat{\Omega}\hat{V}\hat{\Omega}\hat{V}-5\Omega^2\hat{V}^2
+13\textrm{Tr}\{\hat{V}\hat{\Omega}^2\}\hat{V}\\
\hat{W}&=& -2\hat{\Omega}^5-2\hat{V}\hat{\Omega}^3+2\hat{\Omega}^3\hat{V}
+7\hat{\Omega}^2\hat{V}\hat{\Omega}+4\hat{\Omega}\hat{V}\hat{\Omega}^2
+3\hat{\Omega}\hat{V}^2+6\hat{V}\hat{\Omega}\hat{V}+6\hat{V}^2\hat{\Omega}
\end{array}\right.
\end{eqnarray}
\end{subequations}
\end{widetext}
The first constant (\ref{S1}) gives the total energy (Hamiltonian) of the system (\ref{ham}). The remaining constants are higher order expressions in $\hat{V}$ and $\hat{\Omega}$ and they have no direct interpretation. The energy of each mode of the oscillator is a linear combination of these constants.

\end{document}